\documentclass[a4paper,USenglish,cleveref, autoref, thm-restate]{lipics-v2021}

\usepackage{xspace}

\newcommand{\MG}{\textsc{MemGuard}\xspace}
\newcommand{\MP}{\textsc{MemPol}\xspace}

\title{Revisiting MemGuard Overhead: A Reproduction Report} 

\author{Weifan Chen}{Boston University, U.S.}{wfchen@bu.edu}{https://orcid.org/0009-0002-4856-0421}{}

\author{Heechul Yun}{University of Kansas, U.S.}{heechul.yun@ku.edu}{https://orcid.org/0000-0002-8515-2622}{}

\author{Renato Mancuso}{Boston University, U.S.}{rmancuso@bu.edu}{https://orcid.org/0000-0003-3558-5216}{}

\authorrunning{W. Chen, H. Yun, and R. Mancuso} 

\Copyright{Weifan Chen, Heechul Yun, and Renato Mancuso} 



\category{} 

\relatedversion{} 

\nolinenumbers 

\EventEditors{John Q. Open and Joan R. Access}
\EventNoEds{2}
\EventLongTitle{42nd Conference on Very Important Topics (CVIT 2016)}
\EventShortTitle{CVIT 2016}
\EventAcronym{CVIT}
\EventYear{2016}
\EventDate{December 24--27, 2016}
\EventLocation{Little Whinging, United Kingdom}
\EventLogo{}
\SeriesVolume{42}
\ArticleNo{23}

\begin{document}

\maketitle

%
%
%
%
%
%
\begin{abstract}
As an increasing number of embedded platforms incorporate multiple processing units, shared resource contention induced unpredictable execution time poses a challenge for real-time system design. Memory bandwidth regulation is a popular mitigation approach, and MemGuard is the canonical example. Recently, MemPol introduced a new bandwidth regulation mechanism, which was compared with MemGuard. Specifically, they reported significant overheads for MemGuard, citing up to a 1.79x slowdown, to contextualize MemPol's comparative benefits.

This report is meant to clarify and add the necessary nuance to the experiments carried out in that prior work. Specifically, we show that the MemGuard overheads presented in these prior evaluations were unintentionally amplified as the result of using a suboptimal configuration with an older version of MemGuard, wherein the benchmark under test was pinned directly to the master core responsible for handling global timer interrupts. By faithfully reproducing these specific experiments using a modern, decentralized implementation of MemGuard, we demonstrate that the actual execution overhead drops significantly under identical conditions.

Consequently, when evaluated with a properly configured recent version, MemGuard exhibits an overhead that is highly comparable to MemPol's overhead. By revisiting these baseline metrics, this report provides an updated and comprehensive perspective required for future evaluations of memory bandwidth regulators.

\end{abstract}

\section{Introduction}\label{sec:intro}
Multi-core and heterogeneous platforms are now ubiquitous in embedded and real-time systems. In this class of systems, however, the problem of inter-core temporal interference due to the extensive sharing of hardware resources in the memory hierarchy remains an open challenge. Indeed, when multiple cores contend over limited memory resources, memory-access latency becomes highly variable, degrading predictability. Restoring desirable temporal properties in multi-core systems, while avoiding substantial hardware over-provisioning, therefore, requires careful resource management. A key resource that requires management is main memory bandwidth. A widely used software approach to mitigate the problem of uncontrolled memory bandwidth contention is \emph{memory bandwidth regulation}. This encompasses a number of strategies aiming at controlling how much memory traffic a processor can originate within a fixed time window to reduce contention and improve temporal isolation.

The idea of software-based memory bandwidth regulation was first proposed by Bellosa~\cite{bellosa97memoryaccess-tr,bellosa98scheduling_thesis} and later popularized by MemGuard~\cite{yun2013memguard}, which became foundational as the first practical implementation and has since served as a baseline for many subsequent works (e.g., ~\cite{flodin2014dynamic,yun2017bwlock,farshchi2020bru,gifford2021dna,memcore_rtss24,ZBCCM23-mempol,park2025field}).
MemGuard enforces per-core bandwidth budgets by monitoring memory-related events using hardware performance counters and throttling a core once its budget is exhausted; budgets are replenished periodically at a configurable \emph{replenishment period}. In practice, enforcing this periodic replenishment requires timer-driven kernel activity, which inevitably introduces execution-time overhead. Moreover, the replenishment period creates a fundamental tradeoff: shorter periods improve control granularity but trigger more frequent interrupts and typically higher overhead.

Because overhead directly affects both system efficiency and the fairness of comparisons among regulators, it is critical to measure and interpret MemGuard's overhead accurately. This is especially important when overhead numbers are used as a baseline to motivate or evaluate new regulation designs. In particular, the MemPol papers~\cite{ZBCCM23-mempol, ZBCCM24-mempol} report that MemGuard's overhead ranges from $1.79\times$ down to $1.04\times$ as the replenishment period increases from $32\,\mu$s to $1000\,\mu$s, and these results have been used to contextualize MemPol's benefits.

This paper revisits the overhead results reported for MemGuard in MemPol~\cite{ZBCCM23-mempol, ZBCCM24-mempol} through a close collaboration between the authors involved in both line of works, enabling a faithful reproduction of the original setup and a careful examination of MemGuard's implementation evolution.
Specifically, our goal is to reproduce and re-evaluate MemGuard's replenishment-timer overhead under the conditions used in MemPol, and to refine our understanding about how that overhead impacts the MemPol-vs.-MemGuard comparisons.
Our key finding is that the MemGuard overhead reported in~\cite{ZBCCM23-mempol, ZBCCM24-mempol} was significantly inflated (by up to an order of magnitude) because it was obtained using an older MemGuard version that employed a \emph{suboptimal configuration}: the implementation elected a ``master'' core to handle global timer interrupts and send inter-processor interrupts (IPIs), and the benchmark under analysis was pinned to that master core---exactly the situation in which interrupt handling cost is the highest. When this configuration pitfall is avoided (e.g., by pinning the benchmark to a non-master core in the older version) or when using newer MemGuard releases, the measured overhead becomes dramatically smaller (e.g., about $1.09\times$ at $32\,\mu$s).

We further explain why this discrepancy occurs by analyzing MemGuard's internal architectural evolution over the years~\footnote{MemGuard code repository: \url{https://github.com/heechul/memguard}}. In versions prior to commit \texttt{0cf14f2}, MemGuard centralized replenishment-timer handling on a master core and propagated replenishment-period updates via IPIs, creating an asymmetry in interrupt-handling costs across cores; as a result, overhead is abnormally high when the workload runs on the master core. In commit \texttt{0cf14f2} and later, MemGuard switched to a decentralized design where each core maintains its own period using a local timer without global timer/IPI coordination, making timer overhead significantly lower and more uniform across cores.

Finally, using these corrected overhead measurements, we revisit the MemPol experiments that compare MemPol and MemGuard slowdowns (e.g., the evaluations corresponding to Figures~11 and~12 in~\cite{ZBCCM23-mempol, ZBCCM24-mempol}) and discuss how the revised MemGuard overhead changes the interpretation of those results and the apparent advantage of MemPol over MemGuard at comparable regulation periods. Our findings show that with the corrected overhead metric, MemGuard demonstrates highly competitive performance compared to MemPol across all evaluated regulation periods and benchmarks, even outperforming MemPol—especially when regulating memory-intensive benchmarks. In addition, we find that MemGuard's reclaiming and sharing features, which were not evaluated in the original MemPol papers, can significantly improve regulation performance.
\section{Background}

In this section, we provide necessary background on \MG~\cite{yun2013memguard} and MemPol~\cite{ZBCCM23-mempol}. 

\subsection{MemGuard}~\label{sec:mg_overview}

\MG is a software-based memory bandwidth regulation mechanism that leverages performance monitoring counters (PMCs) widely available on modern platforms~\cite{yun2013memguard}. It throttles memory bandwidth by limiting the maximum number of memory transactions a core can perform within a predefined time window. Specifically, each core is assigned a memory budget $Q$, which is consumed as the core performs transactions. This budget is replenished periodically at a fixed interval $P$. If a core depletes its budget before the interval ends, it is throttled until the next replenishment. Consequently, \MG enforces a per-core memory bandwidth limit of $Q/P$. Additionally, \MG supports bandwidth reclamation and sharing mechanisms to improve throughput while maintaining minimum bandwidth guarantees~\cite{yun2013memguard}.

While the original implementation of \MG relied on a single last-level cache (LLC) miss counter to regulate read bandwidth~\cite{yun2013memguard}, it was later extended to support independent read and write regulation. This updated version utilizes LLC refill and LLC write-back counters to track read and write bandwidth, respectively~\cite{bechtel2019dos}. At the beginning of each regulation period, \MG replenishes the budget and programs the performance monitoring unit (PMU) to deliver an overflow interrupt to a core upon budget depletion. When such an interrupt occurs, the core is throttled by scheduling a CPU-intensive, high-priority kernel thread~\cite{yun2013memguard}. Any throttled core is then unblocked at the start of the next regulation period.

Over the years, \MG has undergone a significant architectural change in how it manages regulation periods via timer interrupts. In versions prior to commit \texttt{0cf14f2}, a dedicated master core handled a global period timer interrupt and issued inter-processor interrupts (IPIs) to signal the start of a new period to other cores. This asymmetric design often resulted in higher interrupt handling overhead on the master core. As of commit \texttt{0cf14f2} (released Feb. 22, 2021), this reliance on a global timer and IPIs has been eliminated. Each core now manages its own period using a per-core local timer, ensuring that interrupt handling overhead is distributed uniformly across all cores.

\subsection{MemPol}~\label{sec:mp_overview} 
The following two functionalities of \MG play important roles in memory access regulation: (1) Monitoring the memory bandwidth consumption of a core, and (2) putting a core into a memory-idle state to throttle its memory bandwidth consumption. \MP, also a PMC-based memory bandwidth regulation technique, achieves the said two functionalities employing quite different design and mechanisms compared to \MG. 

For \MP to monitor memory bandwidth, per-core PMCs are accessed from \emph{outside} of the monitored core by an external coprocessor via a memory-mapped interface and periodically polled. 
For example, Xilinx ZYNQ Ultrascale MPSoC (ZCU102) features an Application Processing Unit (APU) cluster and a Real-time Processing Unit (RPU) cluster. \MP can be implemented on (one of) the RPU cores, while user tasks are deployed on the APU cores.
The periodic polling of the PMCs forms a time series of the counter values, which \MP uses to infer the memory bandwidth.

While periodic polling of PMCs for bandwidth monitoring is not new~\cite{yun2012memory}, polling in \MP is different in that it is performed from outside of the monitored cores, by a separate processor. This allows eliminating periodic timer interrupt handling overhead.

However, employing a polling approach has its own downside: the memory bandwidth can \textit{overshoot} the assigned budget. This is because the updated memory bandwidth can only be evaluated at the next polling period. If during the polling period, a burst of memory transactions occurs, the memory bandwidth can already exceed the budget set by \MP before enacting the regulation based on the newly available PMCs readings.

\MG and \MP also differ from how the throttling is achieved. \MP throttles the core activities using the CoreSight debugging interface from ARM. When \MP deems a core need to be throttled, it will instruction the core to enter debug state via the interface, and resume it in a future time.

\subsection{Other Strategies}~\label{sec:other}

More recently, additional alternative approaches have been studied. Most notably, in platforms that include cache-coherent programmable logic (FPGA), it is possible to passively monitor cache snoop requests generated by the last-level cache controller upon a cache miss. This information can be used to accurately compute the per-core memory bandwidth consumption to perform regulation. The approach was demonstrated with the MemCoRe regulator in~\cite{memcore_rtss24}. Furthermore, the ETM$^2$ regulator~\cite{etm2_rtas26} demonstrates the use of Embedded Trace Macrocell (ETM) hardware that is already included in Arm-based processors to implement memory bandwidth regulation. ETM$^2$ relies on passive counting and minimal in-situ processing of microarchitectural events routed to the ETM. This work also includes an additional comparison between \MG and \MP using one of the latest versions of \MG and exploring even shorter regulation periods---down to 20~$\mu s$.

\section{Revisiting the \MG Overhead}\label{sec:setup}

In this section, we revisit the overhead evaluation of \MG in the \MP paper (\cite{ZBCCM23-mempol}; Figure 1), which was used to motivate \MP. 

As discussed in Section~\ref{sec:mg_overview}, \MG uses timer interrupts to handle replenishment periods. As such, replenishment period interrupt handling cost is added overhead that can impact application performance, even when the application never exceeds its memory bandwidth budgets. 

We first define the key overhead metric in this context and then present the experimental results. 
Consider a platform where the benchmark under analysis is the only user task active in the system.
In this context, the overhead is calculated as the slowdown that the benchmark suffers when \MG monitors the core to which the benchmark is pinned but performs no active regulation action. 
This is achieved by setting the allocated bandwidth of the core to an infeasibly large value, so that \MG will only deliver replenishment signals but take no regulation actions. 
Let $C'$ be the execution time of a benchmark under the said condition.
Let $C$ be the baseline execution time of a benchmark. This is the execution time without \MG. The replenishment period induced overhead is defined as $C'/C$.

As in the original MemPol paper, all experiments in this work are conducted on Xilinx Zynq UltraScale+ ZCU102 platform that features four Arm Cortex-A53 application cores with a main memory subsystem comprised of a single DDR4 controller. 
We measured the overhead under various replenishment periods, ranging from 32us to 1ms, using the bandwidth benchmark from the IsolBench~\cite{valsan2016taming}. For a specific setup and replenishment period, ten measurements are conducted.

We conducted the measurement with three different setups: (1) Using the older version of \MG, i.e. the same version as used in \MP, and where the benchmark is pinned to the same core as the main core of \MG; 
(2) Same as (1), but where the benchmark is pinned to another core; 
(3) Using a newer version \texttt{67eb0d8} of \MG . The newer \MG does not elect a main core, thus the host core of the benchmark is randomly selected in each measurement. 



\begin{figure}[h]
    \centering
    \includegraphics[width=0.5\linewidth]{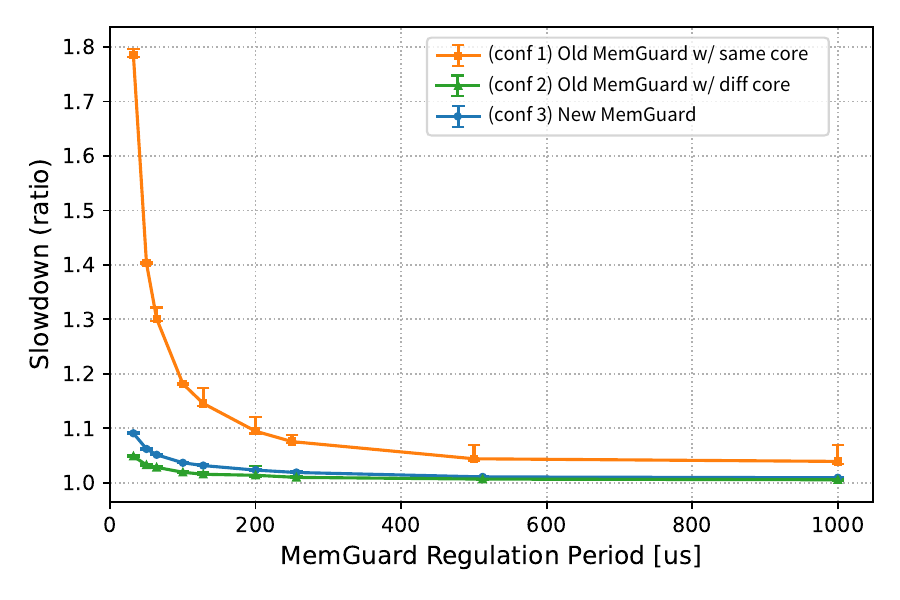}
    \caption{The yellow, blue, and green extrapolated curves show the results of setup (1), (3), and (2) respectively. Each data point is the average of 10 measurements. 
    The error bar indicates the min/max observed value across the 10 measurements. 
    The experiments show that systematically the setup (2) and (3) have significant lower overhead than setup (1). At 32$\mu s$ replenishment period, the slowdowns for setup (1),(2), and (3) are 1.79, 1.05, and 1.09 respectively.}
    \label{fig:slowdown}
\end{figure}

 In \autoref{fig:slowdown}, the blue curve representing the results of setup (3), shows that the overhead is significantly lower than that in setup (1) whose results are represented by the yellow curve. 
The setup (2), represented by the green curve, has the smallest overhead.

The experiments show that the second and third setups have significantly lower overhead than the first setup. At a 32 µs replenishment period, for example, the slowdowns for the first, second, and third setups are 1.79, 1.05, and 1.09 respectively. 
The gaps are narrow in the longer periods but still significant and persistent across all evaluated regulation periods. 

As discussed in Section~\ref{sec:mg_overview}, the first setup is suboptimal because it forces the measured benchmark to run on the master core, which incurs higher overhead due to its centralized design. Using the second (running the benchmark on non-master core in the old MemGuard version) or the third (using newer MemGuard version with de-centralized interrupt handling) configurations, on the other hand, incur dramatically lower overheads.

Notice, the second configuration (old MemGuard with different cores) has even lower overhead than the third configuration (new MemGuard). This is because the old centralizes period timing on a separate main core. The user core receives an inter-process interrupt only to reset and rearm its local PMU counter. The new replaces this with a per-core high-resolution timer, so the user core directly handles the periodic timer interrupt and its control work. Consequently, the old can impose less overhead by offloading global timing work to the main core.

In the following, we will compare the results obtained using the first and third configurations to illustrate how significantly it affects the performance comparison between MemGuard and MemPol.

\section{Re-visiting MemPol and MemGuard Comparison} \label{sec:reeval}
The original \MP paper includes two sets of experiments comparing the slowdown caused by \MP and \MG (Fig.11 and Fig.12 in~\cite{ZBCCM23-mempol}). Because the setup of \MG in these experiments suffered from high overhead, this section presents the re-evaluation of the same experiments with a more up-to-date version of \MG (\texttt{67eb0d8}), and presents a side-by-side comparison. 

The experiments compare the slowdown of the execution time of benchmarks. The slowdown is defined as the ratio of the execution time of the benchmarks under regulation while co-scheduled with other benchmarks, to the unregulated execution time of the benchmarks in isolation.



\begin{figure*}[htp]
    \includegraphics[width=0.95\linewidth]{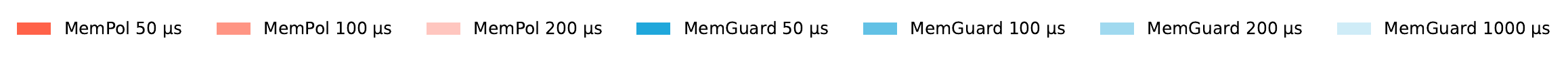}
    \centering
    \includegraphics[width=\linewidth]{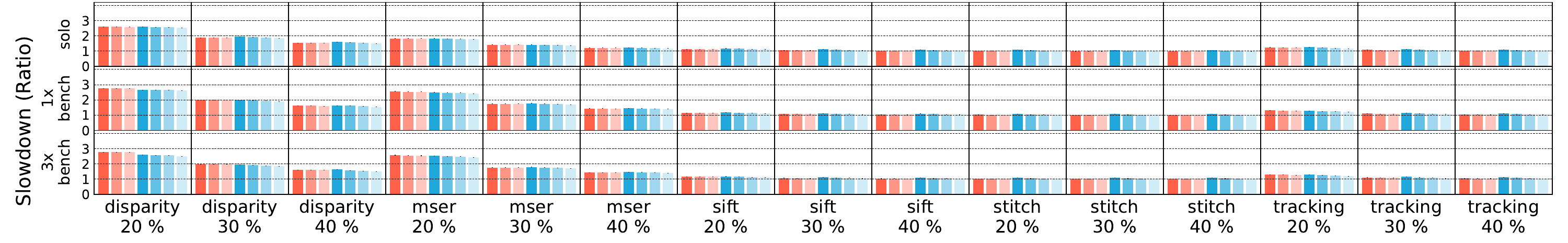}

    \vspace{0.5em}

    \includegraphics[width=\linewidth]{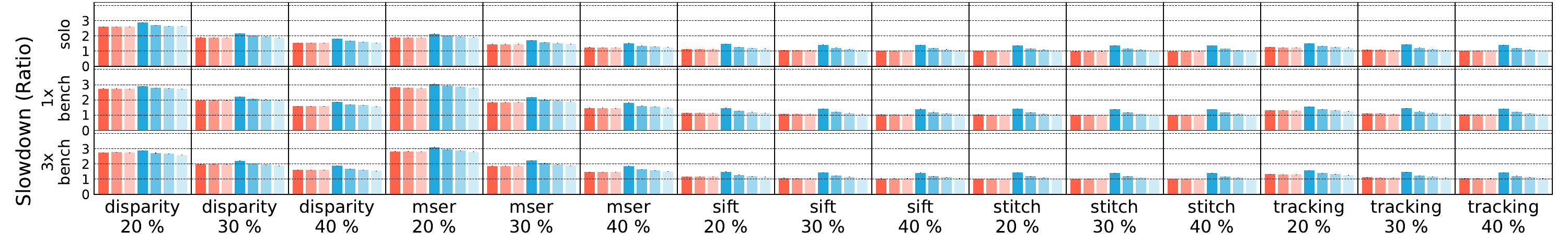}

    \caption{
   The top sub-figure shows the results of modern \MG in which the MemPol data is also newly evaluated, the bottom shows the results from older \MG with overhead abnormality directly taken from the old data. Per sub-figure,
    each row represents the configuration of co-scheduled benchmark(s), while each column represents the benchmark under test at a specific allowed memory bandwidth. Each bar is the average of 10 measurements, with vertical bars indicating the observed min/max overhead.}
    \label{fig:hog}
\end{figure*}

The first experiment uses SD-VBS~\cite{sdvbs} as benchmarks which are regulated at 20\%, 30\%, and 40\% of the sustainable bandwidth, which is empirically estimated to be 1000MB/s on the ZCU102 platform~\cite{Sohal2020EWarPAS} we used for evaluation. 
Either one \textit{bandwidth} benchmark from \textit{IsolBench}~\cite{valsan2016taming} regulated at 60\% of the sustainable bandwidth, or three \textit{bandwidth} each regulated at 20\% of the sustainable bandwidth, are used as co-scheduled benchmarks. All \textit{bandwidth} is conducting memory read access. 
\autoref{fig:hog} presents the results. 

Note first that the bottom sub-figure corresponds to Figure 11 in the original MemPol paper, which suggests that MemGuard suffers from significantly higher overhead compared to MemPol, especially at short regulation periods. In stark contrast, the top sub-figure---obtained using a recent version of MemGuard---shows that MemGuard delivers competitive performance compared to MemPol across all evaluated regulation periods and benchmarks. In fact, for memory-intensive benchmarks such as disparity, MemGuard occasionally outperforms MemPol even at the 50 µs regulation period. This is particularly noteworthy given MemGuard's expected disadvantage due to the higher frequency of timer interrupts at smaller regulation periods.

\begin{figure*}[htp]
    \includegraphics[width=0.95\linewidth]{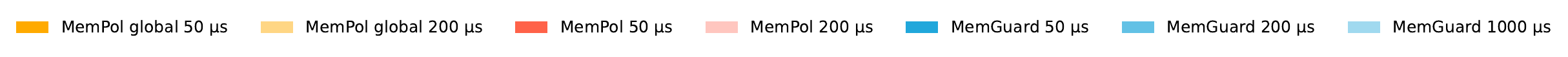}
    \centering
    \includegraphics[width=\linewidth]{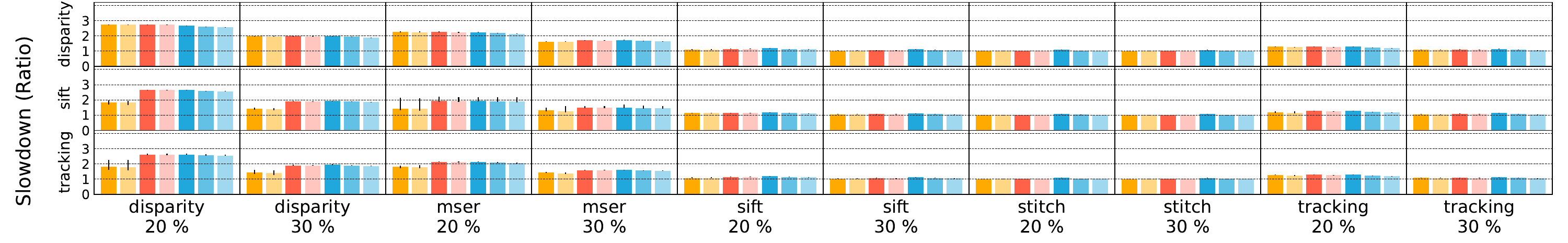}

    \vspace{0.5em}

    \includegraphics[width=\linewidth]{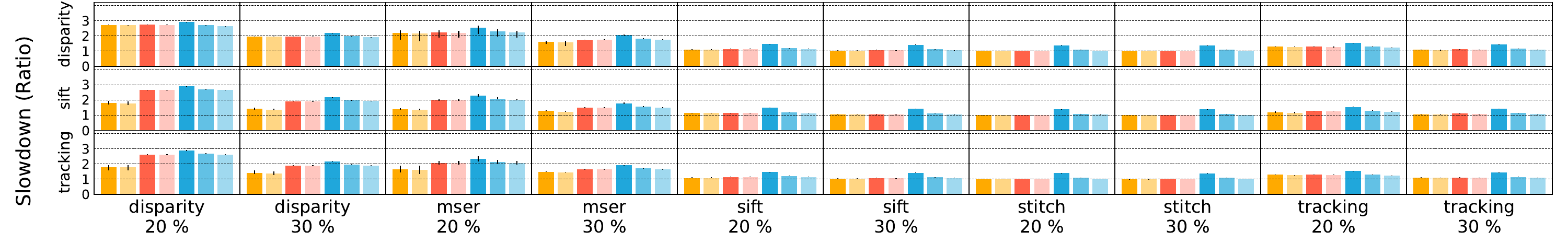}

    \caption{
    The top sub-figure shows the results of modern \MG in which the MemPol data is also newly evaluated, the bottom shows the results from older \MG with overhead abnormality directly taken from the old data. Per sub-figure,
    each row represents the selected co-scheduled contender benchmark. Each column represents the benchmark under test at a specific memory bandwidth. For example, the first column second row represents the overhead of \texttt{disparity} contended by \texttt{sift}. Each bar is the average of 10 measurements with vertical bar indicating the observed min/max.}
    \label{fig:para}
\end{figure*}

The second experiment has two differences compared with the first: (1) \texttt{disparity}, \texttt{sift}, and \texttt{tracking} from SD-VBS are used as co-scheduled benchmarks instead of \textit{IsolBench}. The regulation for both cores are the same; (2) \MP conducts global regulation which allows the allocation of unused global bandwidth (\autoref{sec:mp_overview}). More specifically, the benchmark under test is regulated at 20\% and 30\% of the sustainable bandwidth, with 60\% and 40\% of the global unused bandwidth allocation respectively. \autoref{fig:para} presents the results. 

Consistent with the first experiment, these new results (top sub-figure) demonstrate that modern MemGuard performs comparably to or better than MemPol across all tested benchmarks and periods. This corrects the findings in the original MemPol paper (bottom sub-figure), which utilized a suboptimal configuration. Note that this comparison focuses on basic regulation; MemPol’s global regulation results (first two columns) are discussed in the following section.

\section{Evaluating MemGuard's Bandwidth Reclaiming and Sharing}

This section presents two sets of experiments to evaluate two additional features of \MG: \emph{guaranteed bandwidth reclaiming} and \emph{spare bandwidth sharing}~\cite{yun2013memguard}, against \MP. 
These features can further improve the performance and effectiveness of \MG, but were not evaluated in the original \MP paper~\cite{ZBCCM23-mempol}. This section aims to fill the gap by comparing them with \MP's regulation capability.

In the first experiment, we compare MemPol's \textit{global regulation} with MemGuard's (guaranteed) bandwidth reclaiming feature, as they both aim to re-distribute the sustainable bandwidth from under utilized cores to the needing ones at run-time. 
The basic experiment setup is the same as Figure~\ref{fig:para} except that we include MemGuard with reclaiming enabled. 

\begin{figure*}[htp]
    \includegraphics[width=0.77\linewidth]{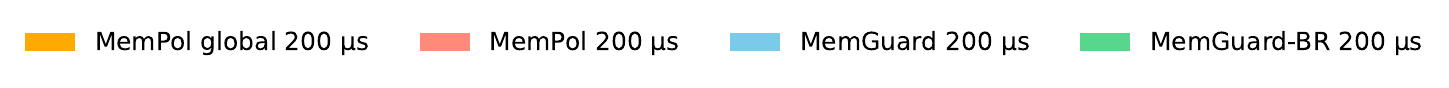}
    \centering
    \includegraphics[width=0.88\linewidth]{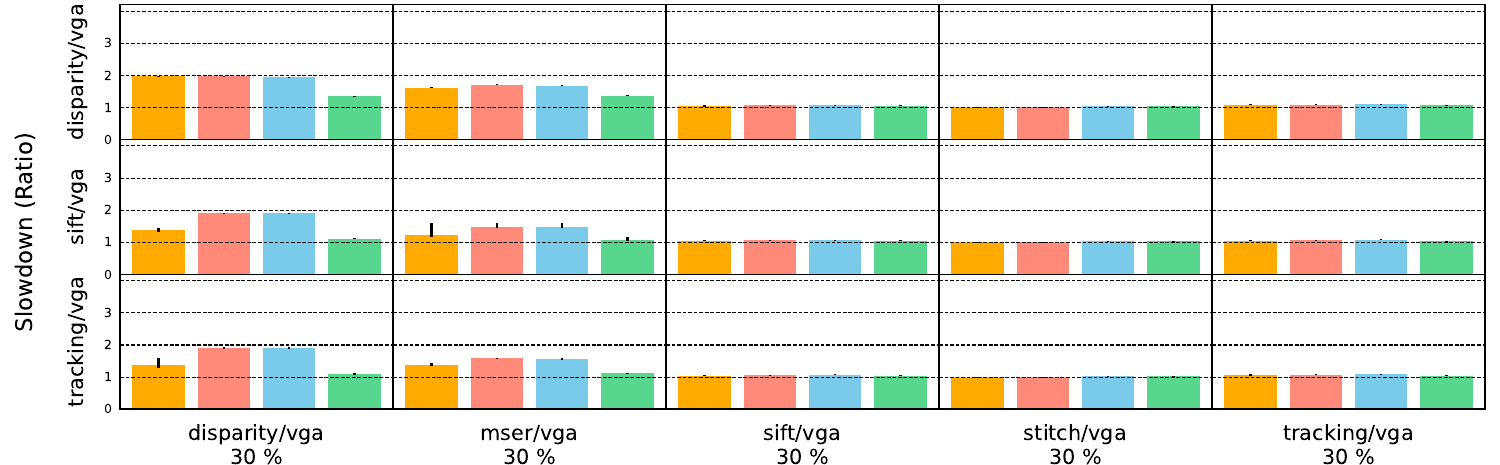}

    \caption{A subset of experiments from \autoref{fig:para}, plus the evaluation of \MG with the \textit{guaranteed bandwidth reclaim} mode (denoted as `MemGuard-BR') enabled.}
    \label{fig:reclaim}
\end{figure*}

Figure~\ref{fig:reclaim} shows the results. With reclaiming enabled, MemGuard can re-distribute reserved but unused bandwidth from the two idle cores in the system to the demanding cores that execute the benchmarks, resulting in significantly improved performance across all tested benchmark combinations. In contrast, MemPol also redistributes unused bandwidth, but its performance gains are less significant than those of MemGuard. 


In the second experiment, we evaluate MemGuard's spare bandwidth sharing feature. 
The experiment setup is as follows: the benchmark is regulated at 20\% of sustainable bandwidth, while 3 co-scheduled \textit{bandwidth} benchmarks are deployed on three other cores, regulated at 20\% each. This is essentially a subset of experiments presented in \autoref{fig:hog}, except that the performance of \MG with spare bandwidth sharing feature enabled is also evaluated.


\begin{figure*}[htp]
    \includegraphics[width=0.77\linewidth]{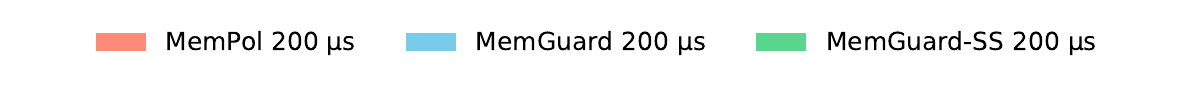}
    \centering
    \includegraphics[width=0.88\linewidth]{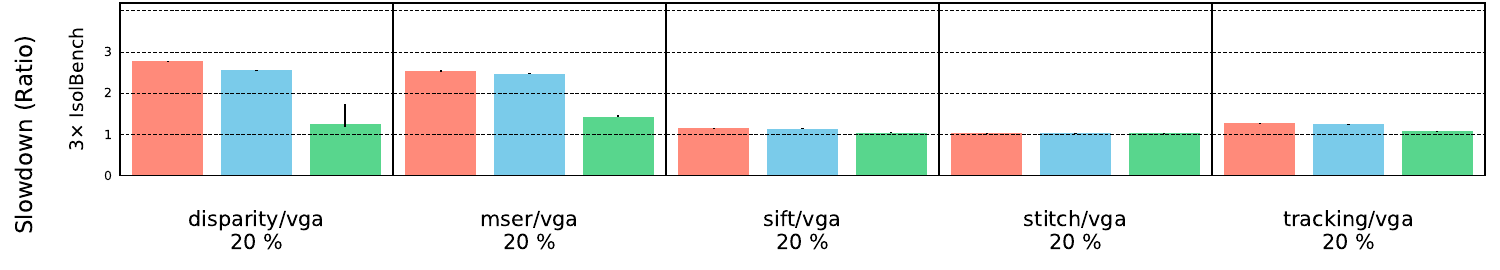}
    \caption{A subset of experiments from \autoref{fig:hog}, plus the evaluation of \MG with the \emph{spare bandwidth sharing} mode (denoted as `MemGuard-SS') enabled.}
    \label{fig:exclusive}
\end{figure*}

\autoref{fig:exclusive} shows the results, which clearly show the spare bandwidth sharing of \MG significantly reduce the slowdown of memory-intensive benchmarks under regulation, such as \texttt{disparity} and \texttt{mser}. This is because within a replenishment period, all the cores quickly deplete their respective budgets. Consequently, \MG allows cores to continue to exploit the spare bandwidth for better throughput. 

Note that, for the spare bandwidth sharing feature, though some experiment results of \MP are presented aside, \MP does not have a comparable feature. Thus the comparison is to illustrate the performance gain of MemGuard's spare bandwidth sharing, not to certain \MP features. Secondly, evaluating \MG's spare bandwidth sharing feature illustrates the potential performance gain if \MP implements similar functionalities.

\section{Conclusion}

This report re-evaluated the execution-time overhead of the MemGuard memory bandwidth regulator to clarify the significantly inflated overheads presented in prior evaluations~\cite{ZBCCM23-mempol,ZBCCM24-mempol}. By faithfully reproducing the original experimental setups, we demonstrated that the previously reported 1.79x slowdown was an artifact of an older, centralized MemGuard implementation where the benchmark under test was pinned directly to the master core responsible for handling global timer interrupts.

When evaluated using a modern, decentralized version of MemGuard---or by simply pinning workloads to non-master cores in the older version---the overhead drops dramatically from 1.79X to roughly 1.09x and 1.05x, respectively. With these corrected baseline metrics, MemGuard demonstrates highly competitive performance compared to MemPol across all evaluated regulation periods. In addition, our supplementary evaluations of MemGuard's guaranteed bandwidth reclaiming and spare bandwidth sharing mechanisms highlight significant performance advantages that further reduce the slowdown of memory-intensive benchmarks. Ultimately, this report provides the updated and comprehensive perspective necessary for fair and accurate future evaluations of memory bandwidth regulators.



\bibliography{references}
\end{document}